\documentclass[letterpaper]{article} 
\usepackage{aaai19}  
\usepackage{times}  
\usepackage{helvet} 
\usepackage{courier}  
\usepackage[hyphens]{url}  
\usepackage{graphicx} 
\usepackage{graphicx}  
\usepackage{subfig}
\usepackage{xspace}
\usepackage{balance}

\usepackage{type1cm}     
\usepackage{graphicx}     
\usepackage{xspace}     
\usepackage{booktabs}     
\usepackage{multirow}      
\usepackage[tableposition=top]{caption}     
\usepackage[vlined,linesnumbered,ruled,noend]{algorithm2e}     
\usepackage{natbib}     
\usepackage{microtype}    
\usepackage{units}     
\usepackage{mathtools}     
\usepackage{amssymb}     
\usepackage{amsmath}
\usepackage{xfrac}    
\usepackage{enumitem}
\usepackage{makecell}
\usepackage{flushend}
\usepackage{mathtools,xparse}
\usepackage{hyperref}
\usepackage[T1]{fontenc} 
\usepackage{lipsum} 
\usepackage{hyphenat}
\usepackage{caption}
\usepackage[most]{tcolorbox}
\usepackage{float} 
\usepackage{mathptmx}
\usepackage{multicol}
\usepackage{xcolor}
\usepackage{wrapfig}

\graphicspath{{img/}}

\def\BibTeX{{\rm B\kern-.05em{\sc i\kern-.025em b}\kern-.08emT\kern-.1667em\lower.7ex\hbox{E}\kern-.125emX}}

\DeclarePairedDelimiter{\norm}{\lVert}{\rVert}

\NewDocumentCommand{\normL}{ s O{} m }{%
    \IfBooleanTF{#1}{\norm*{#3}}{\norm[#2]{#3}}_{F}^2%
}
\NewDocumentCommand{\normF}{ s O{} m }{%
     \IfBooleanTF{#1}{\norm*{#3}}{\norm[#2]{#3}}_{F}%
}
\NewDocumentCommand{\normsquare}{ s O{} m }{%
    \IfBooleanTF{#1}{\norm*{#3}}{\norm[#2]{#3}}^2%
}

\DeclareSymbolFont{greek}{OML}{cmm}{m}{n}
\DeclareMathSymbol{\alpha}{\mathalpha}{greek}{"0B}
\DeclareMathSymbol{\beta}{\mathalpha}{greek}{"0C}
\DeclareMathSymbol{\gamma}{\mathalpha}{greek}{"0D}
\DeclareMathSymbol{\delta}{\mathalpha}{greek}{"0E}
\DeclareMathSymbol{\epsilon}{\mathalpha}{greek}{"0F}
\DeclareMathSymbol{\zeta}{\mathalpha}{greek}{"10}
\DeclareMathSymbol{\eta}{\mathalpha}{greek}{"11}
\DeclareMathSymbol{\theta}{\mathalpha}{greek}{"12}
\DeclareMathSymbol{\iota}{\mathalpha}{greek}{"13}
\DeclareMathSymbol{\kappa}{\mathalpha}{greek}{"14}
\DeclareMathSymbol{\lambda}{\mathalpha}{greek}{"15}
\DeclareMathSymbol{\mu}{\mathalpha}{greek}{"16}
\DeclareMathSymbol{\nu}{\mathalpha}{greek}{"17}
\DeclareMathSymbol{\xi}{\mathalpha}{greek}{"18}
\DeclareMathSymbol{\pi}{\mathalpha}{greek}{"19}
\DeclareMathSymbol{\rho}{\mathalpha}{greek}{"1A}
\DeclareMathSymbol{\sigma}{\mathalpha}{greek}{"1B}
\DeclareMathSymbol{\tau}{\mathalpha}{greek}{"1C}
\DeclareMathSymbol{\upsilon}{\mathalpha}{greek}{"1D}
\DeclareMathSymbol{\phi}{\mathalpha}{greek}{"1E}
\DeclareMathSymbol{\chi}{\mathalpha}{greek}{"1F}
\DeclareMathSymbol{\psi}{\mathalpha}{greek}{"20}
\DeclareMathSymbol{\omega}{\mathalpha}{greek}{"21}
\DeclareMathSymbol{\varepsilon}{\mathalpha}{greek}{"22}
\DeclareMathSymbol{\vartheta}{\mathalpha}{greek}{"23}
\DeclareMathSymbol{\varpi}{\mathalpha}{greek}{"24}
\DeclareMathSymbol{\varrho}{\mathalpha}{greek}{"25}
\DeclareMathSymbol{\varsigma}{\mathalpha}{greek}{"26}
\DeclareMathSymbol{\varphi}{\mathalpha}{greek}{"27}
\DeclareSymbolFont{otone}{OT1}{cmr}{m}{n}
\DeclareMathSymbol{\Gamma}{\mathalpha}{otone}{0}
\DeclareMathSymbol{\Delta}{\mathalpha}{otone}{1}
\DeclareMathSymbol{\Theta}{\mathalpha}{otone}{2}
\DeclareMathSymbol{\Lambda}{\mathalpha}{otone}{3}
\DeclareMathSymbol{\Xi}{\mathalpha}{otone}{4}
\DeclareMathSymbol{\Pi}{\mathalpha}{otone}{5}
\DeclareMathSymbol{\Sigma}{\mathalpha}{otone}{6}
\DeclareMathSymbol{\Upsilon}{\mathalpha}{otone}{7}
\DeclareMathSymbol{\Phi}{\mathalpha}{otone}{8}
\DeclareMathSymbol{\Psi}{\mathalpha}{otone}{9}
\DeclareMathSymbol{\Omega}{\mathalpha}{otone}{10}
\DeclareSymbolFont{syms}{OML}{cmm}{m}{it}
\DeclareMathSymbol{\partial}{\mathord}{syms}{"40}
\DeclareMathAlphabet{\mathbold}{OML}{cmm}{b}{it}
\DeclareSymbolFont{largesymbols}{OMX}{cmex}{m}{n}

\newcommand{\verified}{\textsc{Verified}\xspace}
\newcommand{\political}{\textsc{Political}\xspace}
\newcommand{\despoina}{\textsc{Despoina}\xspace}
\newcommand{\despoinarandom}{\textsc{Despoina (random)}\xspace}
\newcommand{\instagram}{\textsc{Instagram}\xspace}
\newcommand{\youtube}{\textsc{YouTube}\xspace}

\newcommand{\Diff}{\mathrm{d}}
\newcommand{\tmin}{\epsilon}

\newcommand{\hide}[1]{}

\newcommand{\xhdr}[1]{\vspace{1.7mm}\noindent{{\bf #1.}}}

\newcommand{\Secref}[1]{Sec.~\ref{#1}}
\newcommand{\Eqnref}[1]{Eq.~\ref{#1}}

\newcommand{\Figref}[1]{Fig.~\ref{#1}}

\newcommand{\ie}{{i.e.}\xspace}
\newcommand{\eg}{{e.g.}\xspace}
\newcommand{\cf}{{cf.}\xspace}

\newcommand{\etc}{{etc.}\xspace}

\title{Evolution of Retweet Rates in Twitter User Careers: Analysis and Model}

\author{
Kiran Garimella\thanks{Research done while at EPFL.}\\MIT\\garimell@mit.edu
\And
Robert West\\EPFL\\robert.west@epfl.ch
}

\begin{document}
\maketitle


\begin{abstract}
We study the evolution of the number of retweets received by Twitter users over the course of their ``careers'' on the platform. We find that on average the number of retweets received by users tends to increase over time. This is partly expected because users tend to gradually accumulate followers. Normalizing by the number of followers, however, reveals that the relative, per-follower retweet rate tends to be non-monotonic, maximized at a ``peak age'' after which it does not increase, or even decreases. We develop a simple mathematical model of the process behind this phenomenon, which assumes a constantly growing number of followers, each of whom loses interest over time. We show that this model is sufficient to explain the non-monotonic nature of per-follower retweet rates, without any assumptions about the quality of content posted at different times.
\end{abstract}


\section{Introduction}
\label{sec:introduction}

This paper studies the evolution of individual impact in the context of social media.
We focus on Twitter users' \textit{retweet rates}---the number of retweets received by their posts (``tweets'')---as a way to measure the impact and reach of their content. 
We aim to understand how this measure of impact changes over the course of users' ``careers'' (the time since they joined the platform).
Specifically, we ask:
Do users' retweet rates follow certain patterns?
How does a user's retweet rate depend on the size of their audience?
And when in a user's career is their content retweeted most?

Individual impact has been previously studied
in contexts other than social media, including
science~\citep{radicchi2009diffusion,sinatra2016quantifying},
work~\citep{denisi1981profiles,barrick1991big},
sports~\citep{yucesoy2016untangling},
and art~\citep{dennis1966creative,liu2018hot}.
Measuring impact on social media is, however, inherently different, due to the 
highly dynamic nature of online social networks, which may grow (and sometimes shrink) by orders of magnitude within very short time frames, accompanied by numerous confounding factors, including temporal changes in
audience size (number of followers),
productivity (number of posts written),
experience (age on the platform),
interests (content of posts),
and context (external events such elections).
Studying the effects of these factors, and separating them from true impact, is not trivial.
For instance, we observe that on average the impact of individual users, as measured by the number of retweets their content receives, increases over the course of their careers.
Is this due to a real change in content quality, or might it simply be explained by the fact that users accumulate more followers over time?
or due to their increasing expertise in using the platform? or due to external events?

Because of the challenging nature of the problem, research into long-term trends of individual impact on social media, and on the factors that influence these trends, has been limited to date.
Most research has attempted to understand the impact of isolated \textit{pieces of content,} \eg, by predicting retweet counts on Twitter~\citep{suh2010want,figueiredo2011tube,kupavskii2012prediction,martin2016exploring}, video view counts on YouTube~\citep{figueiredo2011tube}, or image like counts on Instagram~\citep{zhang2018become}.
On the contrary, the literature is much thinner regarding the nature and role of the impact of individual \textit{users}.

In order to make progress in this direction, we use an extensive dataset built from Twitter, which spans nearly a decade and contains complete careers of users with a wide range of audience size and experience.
We also obtain historical follower\hyp count estimates of the users in our dataset from the Internet Archive.
Building on this novel dataset, we reconstruct the careers of users---from their very first posts until data collection time---and characterize the ebbing and flowing patterns of individual impact.

Our exposition proceeds in three parts.
In \Secref{sec:retweet_empirical_analysis}, an empirical analysis of absolute and relative (per-follower) retweet rates over the course of Twitter user careers reveals an unexpected non\hyp monotonic progression.
In \Secref{sec:retweet_impact_model}, we propose a mathematical model that captures the crucial non\hyp monotonic aspect of the empirical data while relying on two simple, intuitive assumptions only.
In \Secref{sec:Verification of model assumptions}, we verify that these assumptions are supported by the empirical data.
Finally, in \Secref{sec:discussion}, we discuss our findings and sketch future directions.


\section{Analysis of retweet rates over time}
\label{sec:retweet_empirical_analysis}

\xhdr{Data}
We start from a large dataset spanning almost 10 years (2009--2018) and consisting of all 2.6 billion tweets posted by the more than 600K users who retweeted a U.S.\ presidential or vice\hyp presidential candidate at least five times during that period~\citep{garimella2018political}.
For each user, the dataset contains all tweets they posted.
We focus on each user's original content and discard retweets of others' content.
For each tweet, we also obtained the set of users who retweeted it and the number of times it was retweeted (as of June 2018).
Furthermore, we obtained the set of followers for each user (as of June 2018), as well as time series of historical follower counts estimated via the Internet Archive's Wayback Machine, using Garimella and West's~(\citeyear{garimella2019hot}) method.
Reliable historical estimates of follower counts could be obtained for 23K users, who posted a total of 111M tweets.
Our analyses focus on these 23K users.%
\footnote{
Data available at \url{https://github.com/epfl-dlab/retweet-evolution}
}

\begin{figure}[t]
\centering
\hspace{-4mm}
\begin{minipage}{.47\linewidth}
\subfloat[Retweet count]{
\label{fig:rt_evo_raw}
\includegraphics[width=\textwidth]{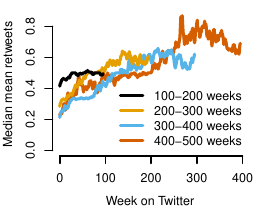}}
\end{minipage}%
\hspace{4mm}
\begin{minipage}{.47\linewidth}
\subfloat[Per-follower retweet rate]{
\label{fig:rt_evo_norm}
\includegraphics[width=\textwidth]{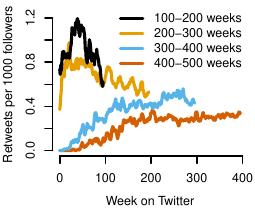}}
\end{minipage}%
\caption{
Evolution of (a)~retweet counts, (b)~per-follower retweet rates in user careers (running averages of width~5).
}
\label{fig:rt_evo}
\end{figure}

\xhdr{Retweet counts over time}
We start with an empirical analysis of the evolution of retweet counts over the course of users' careers.
\Figref{fig:rt_evo_raw} tracks aggregate retweet counts at weekly granularity (where $x=1$ corresponds to a user's first week on Twitter).
To obtain these curves, we first computed, for each user and week, the mean number of retweets obtained by the user for tweets posted that week, and then computed, for each week, the median over all users.
We chose to consider medians of means because retweet\hyp count distributions are heavy-tailed, with a small number of ``viral'' tweets being orders of magnitude more popular than most tweets~\citep{goel2016structural}.
Such outliers can matter a lot for individual users, so to capture them, we consider per-user \textit{means}.
We do not, however, want weekly aggregates to be skewed by outlier tweets, so we consider per-week \textit{medians} of per-user means.

Note that \Figref{fig:rt_evo_raw} contains four curves, each corresponding to a different career\hyp length range (users who have been on Twitter for 100--200 weeks, 200--300 weeks, \etc).
Studying users with different career lengths separately allows us to tease apart the evolution of users from the evolution of the platform and to avoid instances of Simpson's paradox, which might ensue by mixing user groups with systematically different characteristics.

\Figref{fig:rt_evo_raw} shows that the number of retweets obtained per week tends to grow over the course of user careers (with the exception of a final drop for the group of users who have been on Twitter for the longest time).
Moreover, the curves have an overall concave shape, with growth rates slowing down as time progresses.

\xhdr{Per-follower retweet rates}
The growing number of retweets per week may be expected, considering that users tend to accrue followers over time.
To account for this effect, we next consider a normalized version of the retweet rate, where each user's weekly mean number of retweets per tweet is additionally divided by the number of followers the user had at the time.
Aggregating again over all users per week via the median gives rise to \Figref{fig:rt_evo_norm}.
We observe that for some user groups the time series of per-follower retweet rates have a non\hyp monotonic shape with a single peak.
This is particularly the case for users who have been members of Twitter for shorter periods of time (100--300 weeks), whereas for longer-term members (300--500 weeks), the curves grow up to some point, from where on they start to plateau.

Evidence for a ``peak age'' has been found in a wide variety of fields such as poetry, mathematics, and theoretical physics~\citep{adams1946age,dennis1966creative,lehman2017age}.
In line with this literature, one potential narrative for explaining the non\hyp monotonicity could be that users start as ``newbies'', then learn to be good social media users, and finally become old and lose touch with the rest of the community~\citep{danescu2013no}.
In what follows, we will see, however, that such a complex narrative is not necessary to explain the data.
A much simpler model suffices to generate the crucial qualitative shape of the retweets-per-follower curves.


\section{Model of retweet rates over time}
\label{sec:retweet_impact_model}

\begin{figure}[t]
\begin{minipage}{.47\linewidth}
\centering
\subfloat[]{
    \includegraphics[width=\textwidth]{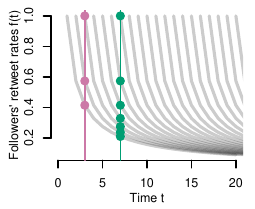}
    \label{fig:modela}
}
\end{minipage}
\begin{minipage}{.47\linewidth}
\centering
\subfloat[]{
    \includegraphics[width=\textwidth]{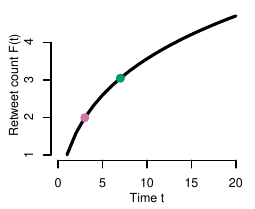}
    \label{fig:modelb}
}
\end{minipage}
\par\medskip
\begin{minipage}{.47\linewidth}
\centering
\subfloat[]{
    \includegraphics[width=\textwidth]{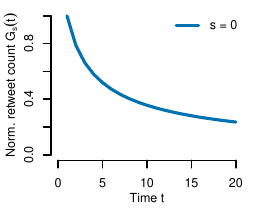}
    \label{fig:modelc}
}
\end{minipage}
\begin{minipage}{.47\linewidth}
\centering
\subfloat[]{
    \includegraphics[width=\textwidth]{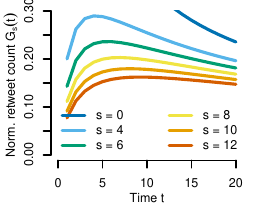}
    \label{fig:modeld}
}
\end{minipage}
\caption{
Model of retweet rates.
(a)~Retweet rates $f(t)$ of individual followers (one curve per follower, shifted along $t$-axis) decay as a power law.
(b)~A user's retweet count $F(t)$ at time $t$ is obtained by summing the retweets from all followers at time $t$.
(c--d)~The relative retweet rate per follower, $G_s(t)$, is obtained by dividing the absolute number of followers by $s+t$, where $s$ is the initial number of followers.}
\label{fig:model}
\end{figure}

Our simple mathematical model makes two assumptions, shown visually in \Figref{fig:modela} and to be verified empirically later (\Secref{sec:Verification of model assumptions}).
\textbf{Assumption~I} states that a user's audience grows at a constant rate, \eg, by one follower per day.
\textbf{Assumption~II} states that each follower's retweet rate $f(t)$ (the expected number of retweets per time unit) decays as a power law:
\begin{equation}
    f(t) = c t^{-\alpha}.
\end{equation}

The number $F(t)$ of retweets at time $t$ is then the sum of the retweets received from all followers accumulated by that time:
$F(t) = \sum_{\tau=1}^t f(\tau)$.
An example is shown in \Figref{fig:modela}:
the magenta and green vertical lines mark specific points in the focal user's career; the intersections of the vertical lines with the followers' retweet-rate curves are indicated by magenta and green dots.
By summing the values of the magenta or the green dots, we obtain the values indicated by the respective dots shown in \Figref{fig:modelb}.
By repeating this procedure for each time step, we obtain the full curve of \Figref{fig:modelb}.
The function $F$ is increasing and concave (\Figref{fig:modelb}), since $f$ is decreasing and positive.
The retweet rate per follower at time $t$ is obtained via $G(t) = F(t)/t$ (without loss of generality, assuming one new follower per time step).
As shown in \Figref{fig:modelc}, $G$ is a monotonically decreasing function.

\begin{figure}[t]
\centering
\hspace{-4mm}
\begin{minipage}{.47\linewidth}
\subfloat[Assumption I]{
\label{fig:assumption1}
\includegraphics[width=\textwidth]{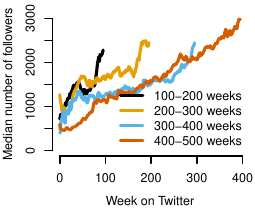}}
\end{minipage}%
\hspace{2mm}
\begin{minipage}{.47\linewidth}
\subfloat[Assumption II]{
\label{fig:assumption2}
\includegraphics[width=\textwidth]{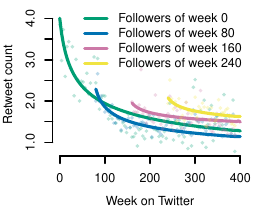}}
\end{minipage}%
\caption{Empirical validation of model assumptions.} 
\label{fig:assumptions}
\end{figure}

Empirical analogues of $F$ and $G$ were displayed in \Figref{fig:rt_evo}.
While the shape of the theoretical $F$ (\Figref{fig:modelb}) is approximately mirrored by empirical data (\Figref{fig:rt_evo_raw}),
there is a discrepancy for $G$, which is monotonically decreasing in the model (\Figref{fig:modelc}), but not so in empirical data (\Figref{fig:rt_evo_norm}).
The discrepancy is resolved by letting users start not with zero but with $s>0$ followers. As we will see later (\Figref{fig:assumption1}), this is in line with empirical data.
An initial follower count $s>0$ could potentially be caused by the platform recommending newcomers to other users more aggressively right after joining, so they would quickly accumulate a certain number of followers immediately in the beginning.
With $s \geq 0$ initial followers,
the relative number of retweets per follower is
$G_s(t) = F(t)/(s+t)$,
which has an internal maximum for $s>0$ (\Figref{fig:modeld}) and captures the essence of the empirical curves of \Figref{fig:rt_evo_norm}.

The curves shown in \Figref{fig:model}a--d were obtained empirically for a specific set of parameters ($\alpha=0.8, c=1$).
In the general case, the model is more easily analyzed when assuming continuous, rather than discrete, time:

\begin{eqnarray}
F(t) &=& \int_0^t f(\tau+\tmin) \;\Diff \tau  \label{eqn:F} \\
    &=&
    \int_0^t c(\tau+\tmin)^{-\alpha} \;\Diff \tau \\
     &=& \frac{c}{\alpha-1} \left( \tmin^{-(\alpha-1)} - (t + \tmin)^{-(\alpha-1) } \right),
\end{eqnarray}
for $\alpha \neq 1$.%
\footnote{
\label{footnote:alpha=1}
The case $\alpha=1$ needs to be handled separately:
here, $F(t) = c \left( \log(1 + t/\tmin) \right)$.
}
The offset $\tmin > 0$ is necessary to be able to handle the case $\alpha \geq 1$, where
$\int_0^t f(\tau) \;\Diff \tau = \int_0^t c \tau^{-\alpha} \;\Diff \tau = \infty$ would diverge.

In the continuous case, $F$ has the same increasing concave shape as in the discrete case (\Figref{fig:modelb}).
The shape of $G_s$ (the retweets per follower over time) is analyzed in detail in Appendix~\ref{sec:proof},%
\footnote{
Appendices available online at \url{http://arxiv.org/abs/2103.10754}
\label{fn:appendix}
}
where we show that $G_s$ attains a unique internal maximum at a time $t^* > 0$ if and only if $s>0$ (as in the discrete case; \Figref{fig:modeld}).

To summarize, the non\hyp monotonic curves indicating a ``peak age'' on Twitter (\Figref{fig:rt_evo_norm}) do not require a complex narrative as sketched earlier (inexperienced youth, age of maximal alignment with community norms, losing touch with newer members in old age).
Rather, they can be explained with a much simpler model containing just two ingredients: (1)~a constant influx of new followers, (2)~each of whom loses interest over time according to a power law.

To be clear, we do not claim that our model fully captures all the dynamics of retweet rates. Rather, we have shown that the seemingly complex, non\hyp monotonic evolution of retweets per follower directly follows from two intuitive assumptions, which are shown to approximately hold in empirical data in the following section.

As a side note, we remark that, for $\alpha>1$, $F(t)$, the number of retweets per time unit, is upper-bounded by a constant
$\frac{c \tmin^{-(\alpha-1)}}{\alpha-1}$.
That is, for $\alpha>1$, the model predicts an inherent barrier on the number of retweets a user can get in a single time unit, even if her audience grows at a constant rate: the drop-off of interest would be too steep to be compensated by the influx of new followers.
Later, we will observe empirically (\Figref{fig:powerlaw_params}) that the empirical exponents $\alpha$ are well below~1, which theoretically implies a potentially unbounded number of retweets per time unit.


\section{Validation of model assumptions}
\label{sec:Verification of model assumptions}

We now verify that the model's assumptions hold empirically.

\xhdr{Assumption I: Constant audience growth}
\Figref{fig:assumption1} shows that the median number of followers indeed grows approximately at a constant rate.
We also see that, empirically, users tend to start off with a nonzero number $s>0$ of followers, which is the condition for which the per\hyp follower retweet rate $G_s(t)$ is non\hyp monotonic in our model (\Figref{fig:modeld}).
Non\hyp monotonic empirical curves (\Figref{fig:rt_evo_norm}) are thus in line with the shape predicted by the model.

\begin{figure}
\begin{center}
\includegraphics{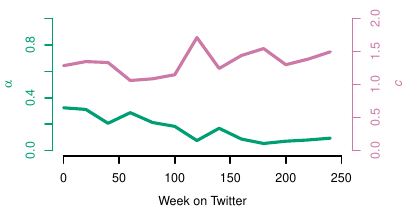}
\end{center}
\caption{
Empirically estimated parameters (power-law exponent $\alpha$ and multiplicative factor $c$) of followers' retweet rates $f(t)=ct^{-\alpha}$, as functions of the week in the focal user's career when the followers began to follow the focal user.
}
\label{fig:powerlaw_params}
\end{figure}

\xhdr{Assumption II: Decay of followers' interest}
To verify the assumption that each follower's retweet rate $f(t)$ decays as a power law, we proceed as follows.
For each week $t_0$ in a user $u$'s career, consider all followers who retweeted $u$ for the first time that week and track them for the rest of $u$'s career, computing these followers' mean numbers of retweets of $u$'s tweets for each subsequent week $t' \geq t_0$, and aggregating weekly over all users $u$ via the median.
The result is shown in \Figref{fig:assumption2} (for users who have been on Twitter for at least 400 weeks; for clarity's sake, curves are shown for only four starting weeks $t_0$, but in practice we computed curves for all $t_0$).
We observe that decreasing power laws $ct^{-\alpha}$ (solid lines in \Figref{fig:assumption2}) fit followers' interest well. 
Irrespective of when a follower starts retweeting, their interest in the followed user drops off with time.
\Figref{fig:powerlaw_params}, which tracks $\alpha$ and $c$ over time, shows that the drop-off in interest becomes less steep over time (as indicated by the decreasing values of $\alpha$).


\section{Discussion}
\label{sec:discussion}

\xhdr{Summary}
Our empirical analysis (\Secref{sec:retweet_empirical_analysis}) of the ``careers'' of thousands of Twitter users, starting with each user's very first tweet, revealed that the number of retweets a user receives per week grows steadily, at a decreasing speed (\Figref{fig:rt_evo_raw}).
The growth is partly due users' tendency to accrue followers over time (\Figref{fig:assumption1});
the decreasing speed, due to followers' tendency to lose interest over time (\Figref{fig:assumption2}).

Accounting for the growing number of followers by dividing weekly retweet counts by the number of followers at the given time reveals partly non\hyp monotonic curves of retweets per follower, especially for more recent users (\Figref{fig:rt_evo_norm}).
One might be tempted to explain this shape by positing a prototypical user life cycle, where users start off unacquainted with the norms of the platform, subsequently produce content that is gradually becoming more attractive to others, until they reach a peak age after which they cease to be ``in tune'' with the rest of the community, resulting in decreasing retweet rates per follower from there on.

By formulating a mathematical model (\Secref{sec:retweet_impact_model}), we show that such a complex narrative is not required in order to explain the non\hyp monotonic shape of retweets-per-follower curves.
Rather, we can reproduce the qualitative shape of the curves by making only two simple, intuitive assumptions, both of which were verified to hold empirically in the data (\Secref{sec:Verification of model assumptions}).
The assumptions state that
(1)~users gain followers at a steady rate, and
(2)~the followers lose interest in the followed user's content according to a power law.
We emphasize that it is not our goal to prove that the model captures all the dynamics that drive retweet rates---in fact, there are probably many factors and facets that remain unmodeled.
Rather, we show that the simple model is sufficient to explain key aspects of the data.

\xhdr{Further Twitter datasets}
The above analyses were performed on a Twitter dataset that is particularly well suited for studying entire user careers, as it contains all tweets posted by
a predetermined set of users.
Collecting this dataset was demanding, as it required custom\hyp made crawling tools, due to the fact that Twitter's API only provides access to each user's 3,200 most recent tweets \citep{garimella2019hot}.
Although our user sample is not random, but biased towards politically interested users \citep{garimella2018political}, we consider this topical bias to be less severe than the bias that would be induced by studying only users with fewer than 3,200 tweets, the limit imposed by Twitter's API.
Nevertheless, to further corroborate our findings, we ran our analyses on three other Twitter datasets \citep{garimella2019hot}, which are not topically biased, but are subjected to the 3,200-tweet limit.
The results, reported in Appendix~\ref{sec:Additional datasets} (\cf\ footnote~\ref{fn:appendix}),
are qualitatively coherent to those reported here.

\xhdr{Beyond Twitter}
Going even further, we collected and analyzed full user careers from two other platforms, Instagram and YouTube.
Although these platforms are rather different from Twitter, we found patterns for absolute and per-follower like counts on Instagram and view counts on YouTube that mirror the patterns for retweet counts found on Twitter.
The results are presented in Appendix~\ref{sec:Additional datasets}
(\cf\ footnote~\ref{fn:appendix}).

\xhdr{Future work}
To the best of our knowledge, this is the first longitudinal study of complete user careers on social media. 
Our findings, public dataset, and models can serve as a stepping stone for further research in this area.
For instance, whereas here we only considered the \textit{number} of followers, one could also use heuristics~\citep{meeder2011we} to approximate the actual \textit{set} of followers.
This could allow us to study how the follower network evolves and how it influences, and is influenced by, retweeting behavior \citep{myers2014bursty}.
Moreover, it would be interesting to define scores for measuring the impact of social media users, similar to the scores used to measure the impact of scientists~\citep{sinatra2016quantifying}.
From a practical perspective, it would be useful to infer long-term behavioral patterns associated with successful user careers.
For instance, can the peak age be delayed, or the drop-off be slowed down, by acting in certain ways?
Beyond science alone, the answers to such questions would be of value for users, marketers, and social media platforms striving for better user retention.



\vfill
\section*{Acknowledgments}
This work was supported in part
by grants from
the Swiss National Science Foundation and
the Swiss Data Science Center,
and by gifts from
Google, Facebook, and Microsoft.

\appendix
\section{Analysis of model properties}
\label{sec:proof}

In \Secref{sec:retweet_impact_model} we claimed that, for any $s$, the shape of the retweet-count-per-follower curve $G_s(t)$ predicted by the model resembles that of the example curves in \Figref{fig:modeld}.
To prove this for the general case, we will show that $G_s(t)$ has a \textbf{unique extremum} $t^*$, which is a \textbf{maximum} and which is \textbf{internal} (i.e., $t^* > 0$) if and only if $s>0$.

For convenience, we first repeat the definitions of
the retweet rate $f(t)$ (the expected number of a follower's retweets per time unit; identical for all followers),
the cumulative number $F(t)$ of retweets gathered from all followers by time $t$,
and the relative number $G_s(t)$ of retweets per follower when assuming $s \geq 0$ initial followers:%
\footnote{As mentioned, the model is more readily analyzed when assuming continuous, rather than discrete, time, so we work with continuous time here.}
\begin{align}
f(t) &= c t^{-\alpha} \\
F(t) &= \frac{c}{\alpha-1} \left( \tmin^{-(\alpha-1)} - (t + \tmin)^{-(\alpha-1) } \right) \\
G_s(t) &= \frac{F(t)}{s+t}
\end{align}

The above definitions assume $\alpha > 0$ and $\alpha \neq 1$ (\cf\ footnote~\ref{footnote:alpha=1}).

Using $\phi'$ to denote the derivative of a function $\phi$, we find the extrema of $G_s(t)$ by setting its derivative to zero and using the fact that $F'(t) = f(t + \tmin)$ (\Eqnref{eqn:F}):
\begin{align}
    G_s'(t) &= 0 \label{eqn:extremum condition} \\
    \frac{F'(t)}{s+t} - \frac{F(t)}{(s+t)^2} &= 0 \\
    f(t+\tmin) \; (s+t) - F(t) &= 0 \\
    c \, \frac{s+t}{(t+\tmin)^{\alpha}} - \frac{c}{\alpha-1} \left( \tmin^{-(\alpha-1)} - (t + \tmin)^{-(\alpha-1)} \right) &= 0 \\
    (\alpha-1) \, (s+t) - \tmin^{-(\alpha-1)} (t + \tmin)^\alpha + t+\tmin &= 0 \\
    \underbrace{\tmin^{-(\alpha-1)} (t + \tmin)^\alpha - (\alpha-1) \, s - \tmin}_{h(t)} &=  \alpha t \label{eqn:line vs h}
\end{align}

Note that, since $\alpha > 0$ and $t \geq 0$, the function $h(t)$ is monotonically increasing.
Moreover, $h(t)$ is convex for $\alpha > 1$, and concave for $\alpha < 1$.
Finally, the slopes of the left- and right-hand sides in \Eqnref{eqn:line vs h} are equal when
\begin{align}
    h'(t) &= \alpha \\
    \alpha \, \tmin^{-(\alpha-1)} (t + \tmin)^{\alpha-1} &= \alpha \\
    t &= 0.
\end{align}

It follows that, since $h(t)$ is monotonically increasing and either convex or concave, $\alpha t$ and $h(t)$ can intersect at most once; \ie, $G_s(t)$ has at most one extremum $t^*$.
If $h(t)$ is convex [concave], an intersection (and thus an extremum of $G_s(t)$) exists if and only if $h(0) \leq 0$ [$h(0) \geq 0$].
Since, as stated above, $h(t)$ is convex if $\alpha > 1$, and concave if $\alpha < 1$, $G_s(t)$ has an extremum if and only if
\begin{align}
    h(0) \, (\alpha-1) &\leq 0 \\
    - (\alpha-1)^2 \, s &\leq 0 \\
    s &\geq 0,
\end{align}
which is true by definition, so an extremum always exists.
In combination with the above fact that there is at most one extremum, this implies that, for any $s \geq 0$, $G_s(t)$ has a \textbf{unique extremum.}
The extremum is an \textbf{internal extremum} (\ie, $t^* > 0$) if additionally $h(0) \neq 0$, \ie, if $s>0$.

Finally, by consistently replacing the equality signs with inequality signs in \Eqnref{eqn:extremum condition}--\ref{eqn:line vs h}, it can be seen that,
for $\alpha > 1$ (\ie, for convex $h$),
$G_s'(t) > 0$ if and only if $h(t) < \alpha t$,
and for $\alpha < 1$ (\ie, for concave $h$),
$G_s'(t) > 0$ if and only if $h(t) > \alpha t$.
Since $h(t)$ is monotonically increasing,
this implies that 
$G_s'(t) > 0$ if and only if $t < t^*$.
Analogously, $G_s'(t) < 0$ if and only if $t > t^*$.
That is, $G_s(t)$ is increasing to the left of $t^*$, and decreasing to the right of $t^*$, and it follows that the \textbf{unique internal extremum is a maximum.}



\section{Additional datasets}
\label{sec:Additional datasets}

Whereas the main part of the paper focused on one particular sample of tweets (\cf\ \Secref{sec:retweet_empirical_analysis}),
for robustness we also ran the analysis on five further datasets, three from Twitter, one from Instagram, and one from YouTube.
In this appendix, we first briefly describe each of the six datasets (including the one from the main part of the paper, henceforth called \political) and then reproduce the key empirical results of \Figref{fig:rt_evo} for all six datasets.

\subsection{Description of datasets}
\label{sec:Description of datasets}

\xhdr{\political}
The dataset from the main part of the paper contains around 2.6 billion tweets from over 600K politically interested users spanning a period of almost 10 years (2009--2018). The dataset consists of \textit{all} tweets posted by users who retweeted a U.S.\ presidential or vice-presidential candidate from 2008 to 2016 at least five times.
In order to obtain all tweets by a user (rather than just their 3,200 most recent tweets, the limit imposed by the Twitter API), a Web-scraping tool,%
\footnote{\url{https://github.com/Jefferson-Henrique/GetOldTweets-python}}
rather than the Twitter API, was used.

\xhdr{\verified}
Starting from a list of all 297K verified Twitter users as of June 2018,%
\footnote{\url{http://redd.it/8s6nqz} (last visited 14 January 2020)}
we maintained only users with between 2,000 and 3,200 tweets and between 50K and 2M followers.
The upper bound on the number of followers was necessary due to the strict rate limiting constraints on the Twitter API endpoint for getting followers.
The upper bound on the number of tweets was imposed in order to ensure that we could capture complete user careers, given that the Twitter API provides only the 3,200 most recent tweets of a user.

\xhdr{\despoinarandom}
We randomly sampled 1.2M users with at most 3,200 tweets from a large sample of 93M Twitter users \citep{antonakaki2018utilizing} and downloaded all their tweets using the Twitter API.

\xhdr{\despoina}
Again starting from the 93M users of~\citet{antonakaki2018utilizing}, here we first constrained the data on certain criteria before sampling randomly.
We only selected users who
(1)~posted between 2,000 and 3,200 tweets (to study only reasonably active users),
(2)~created their account at least six months before April 2016 (the time when this dataset was collected),
and (3)~had at least 100 followers.
From the remaining 6M users, we randomly sampled 1M users and included all their tweets.

\xhdr{\instagram}
Starting from a sample of 50K Instagram users collected by \citet{vilenchik2018million}, we selected the users for whom we could locate historical data on the Internet Archive, such that we could estimate historical follower counts as described in \Secref{sec:retweet_empirical_analysis}.
This resulted in about 4,000 accounts, for each of which we obtained the images they shared and the corresponding like counts.
In our analysis of Instagram, shared images take the role played by tweets in our analysis of Twitter.

\xhdr{\youtube}
We randomly sampled 50K channels from a list of channels obtained from \citet{abisheva2014watches}. We only retained channels with between 100 and 3,000 videos and obtained the view counts for all videos in a channel via the YouTube API.
In our analysis of YouTube, videos take the role played by tweets in our analysis of Twitter, and views take the role played by retweets.

\vspace{2mm}
\noindent
Note that, for the three Twitter datasets other than \political, user sampling is biased:
for a fixed level of user activity (\eg, in terms of tweets posted per day), the older a user is, the more likely they are to have been excluded because of the 3,200-tweet limit imposed by the Twitter API, such that the samples might contain less-active users in older than in younger age groups.
Partly, this concern is mitigated by stratifying the analysis by career length (100--200 weeks, 200--300 weeks, \etc).
More importantly, however, we emphasize that the dataset in the focus of the main part of the paper, \political, is not affected by the 3,200-tweet limit imposed by the Twitter API and does hence not systematically bias older career groups to contain less-active users.


\balance

\vfill
\subsection{Results}
\label{sec:Results on additional datasets}

\begin{figure*}
\centering
\includegraphics[width=0.33\textwidth]{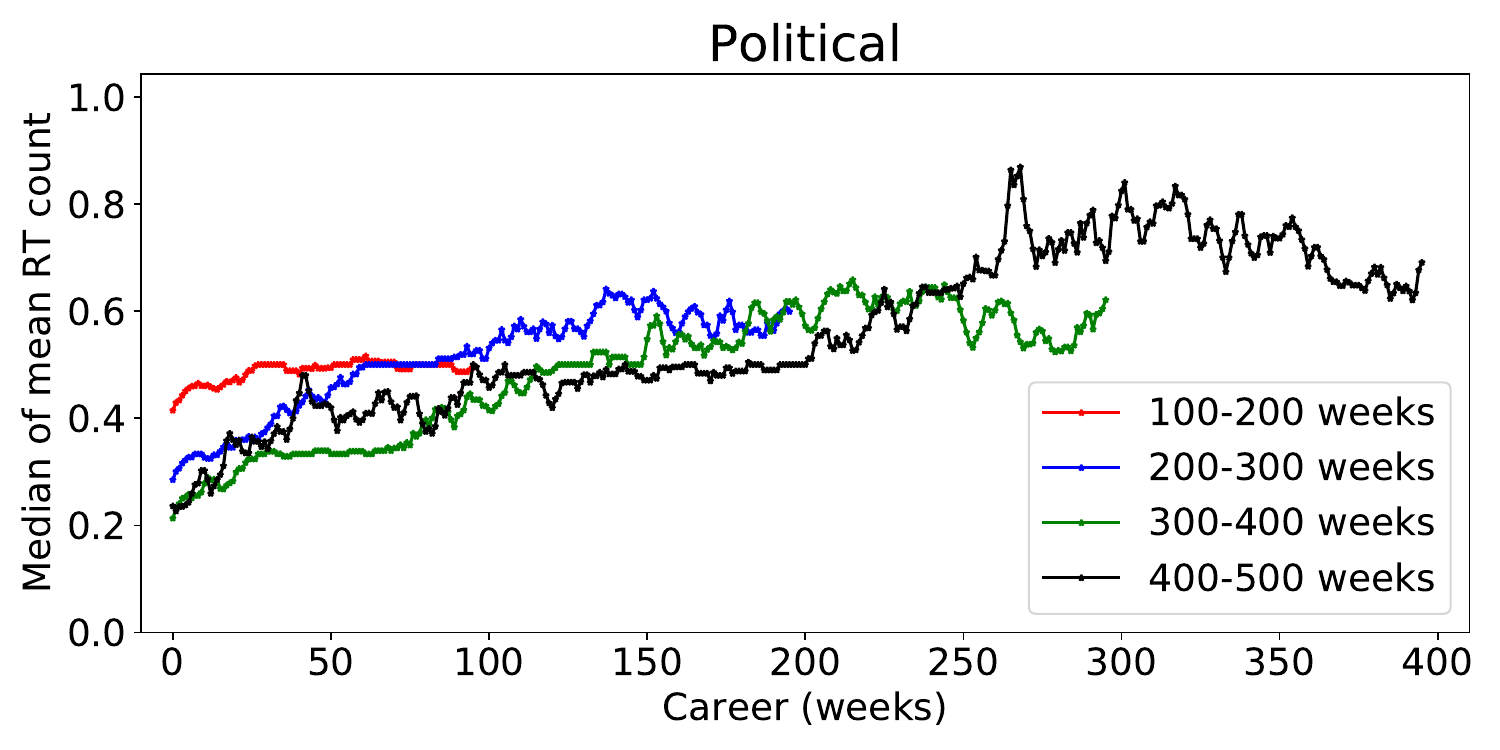}
\includegraphics[width=0.33\textwidth]{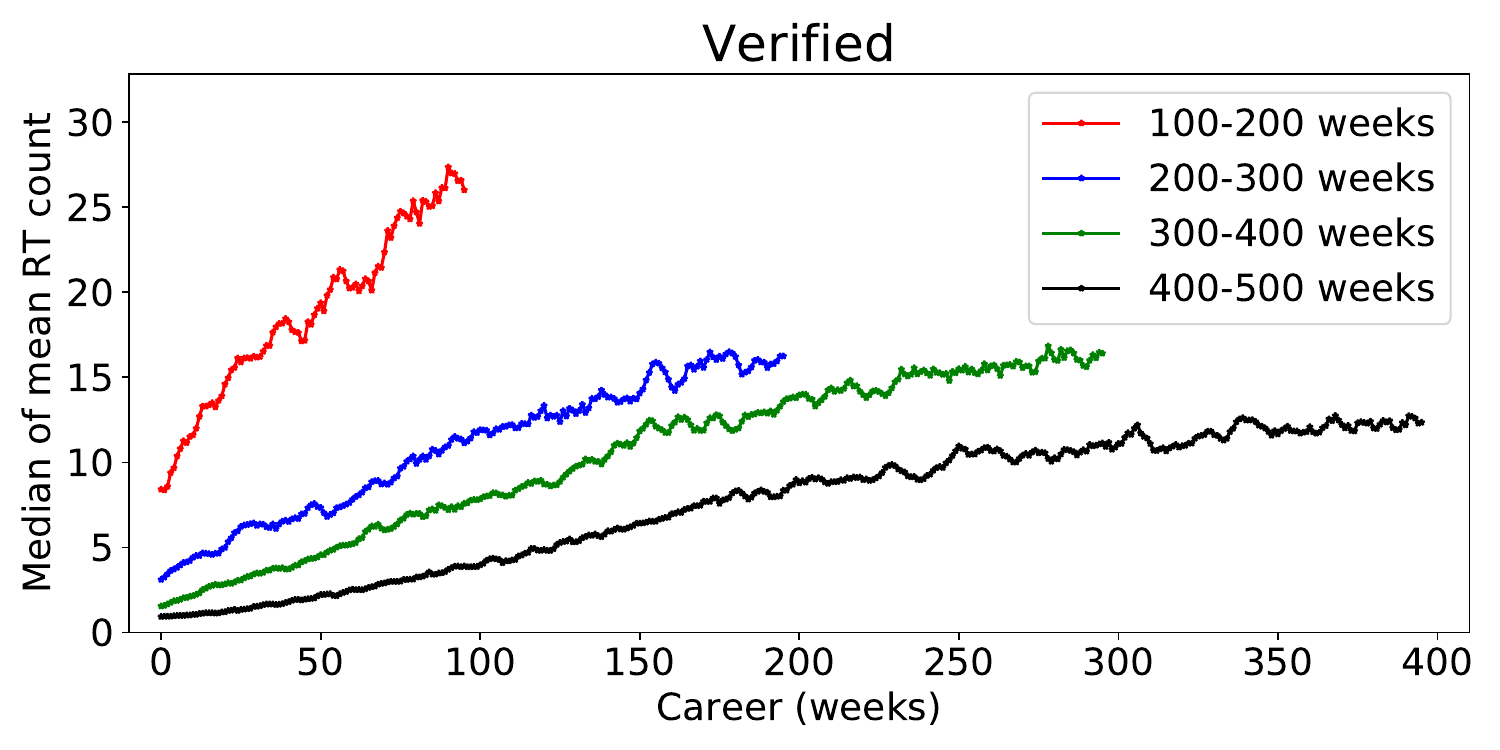}
\includegraphics[width=0.33\textwidth]{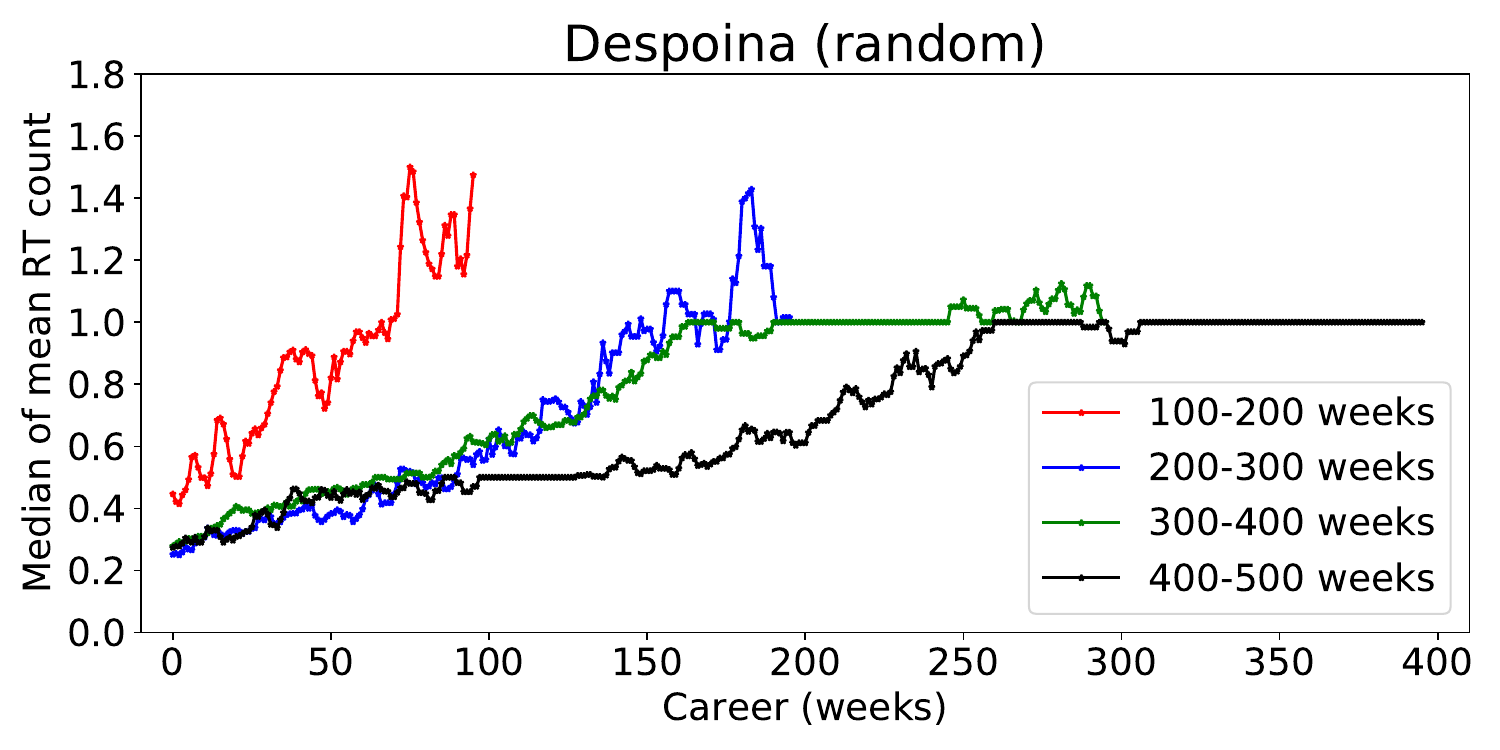}
\includegraphics[width=0.33\textwidth]{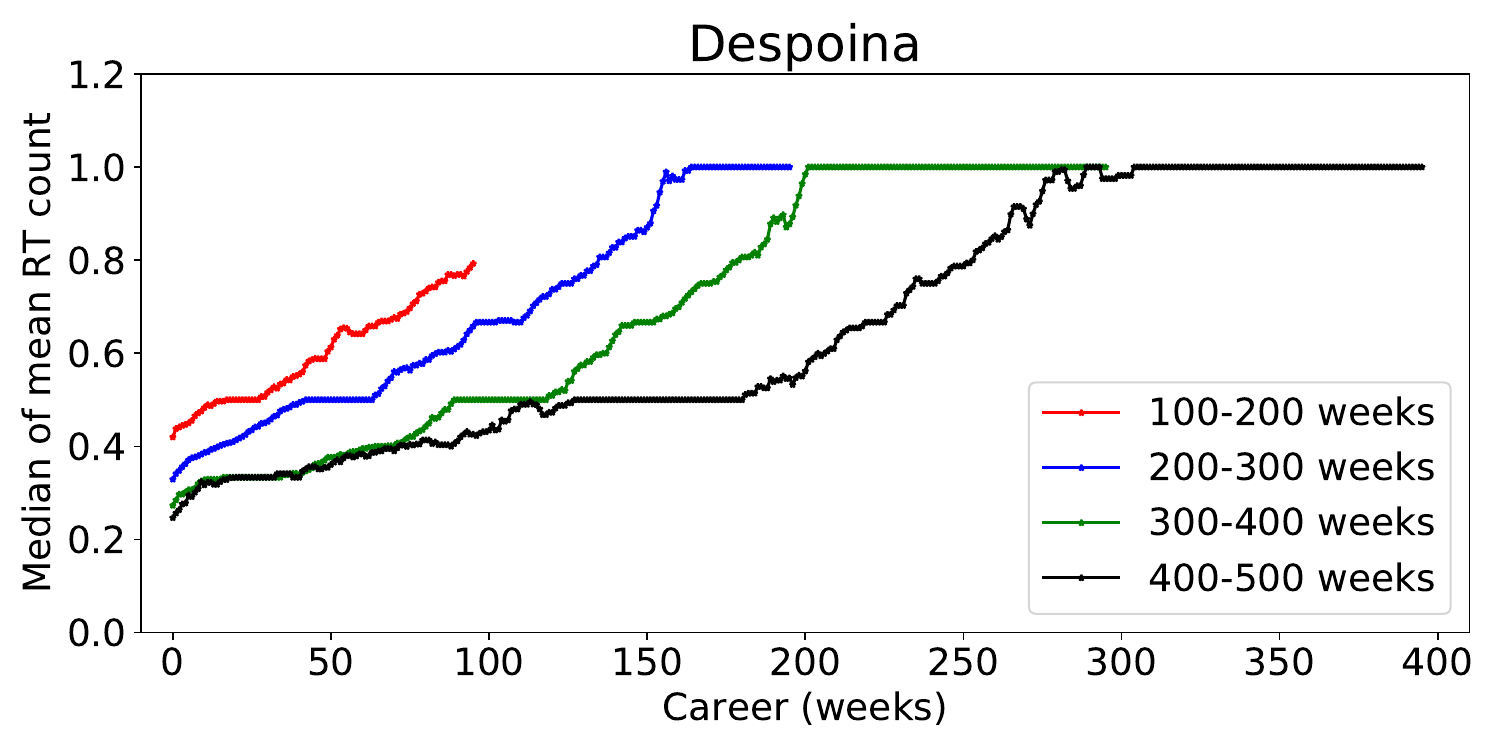}
\includegraphics[width=0.33\textwidth]{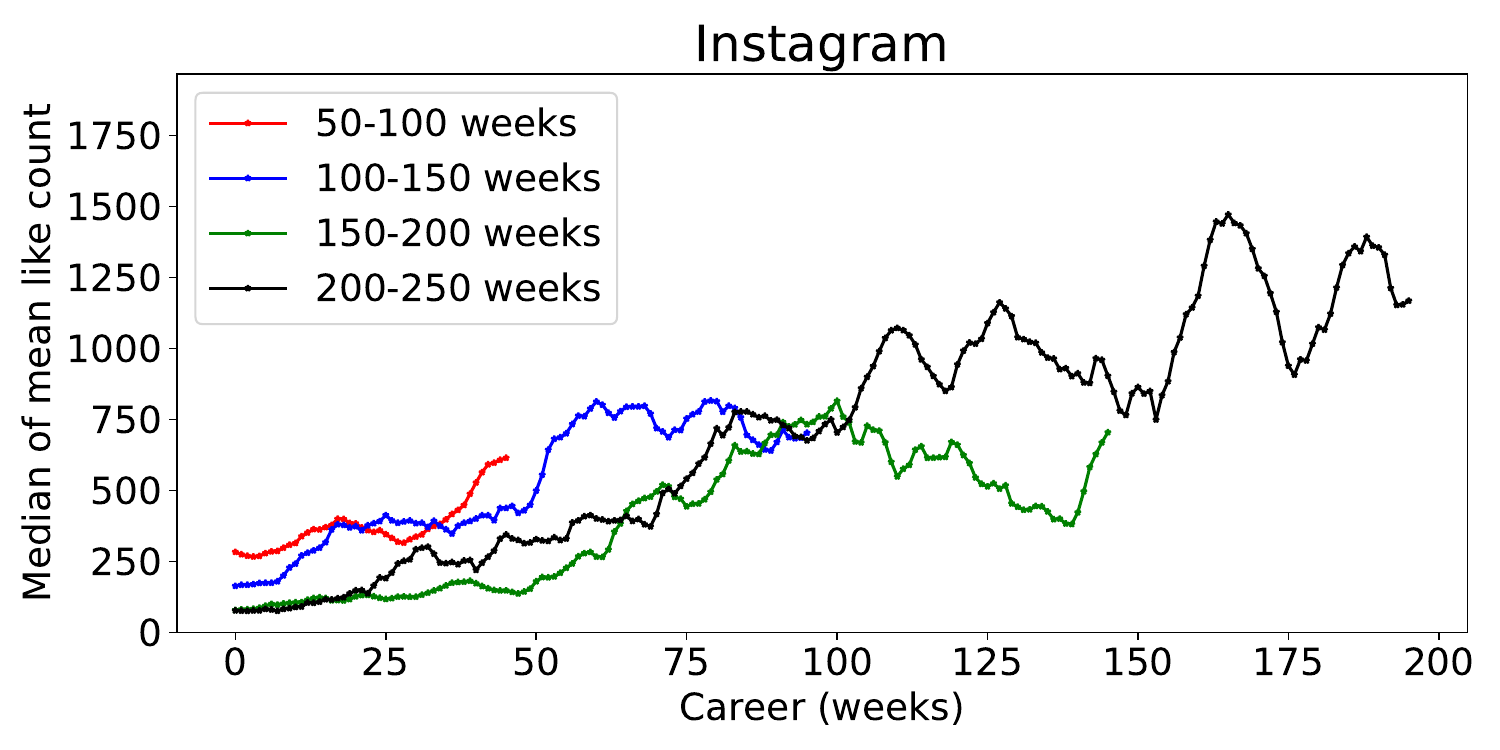}
\includegraphics[width=0.33\textwidth]{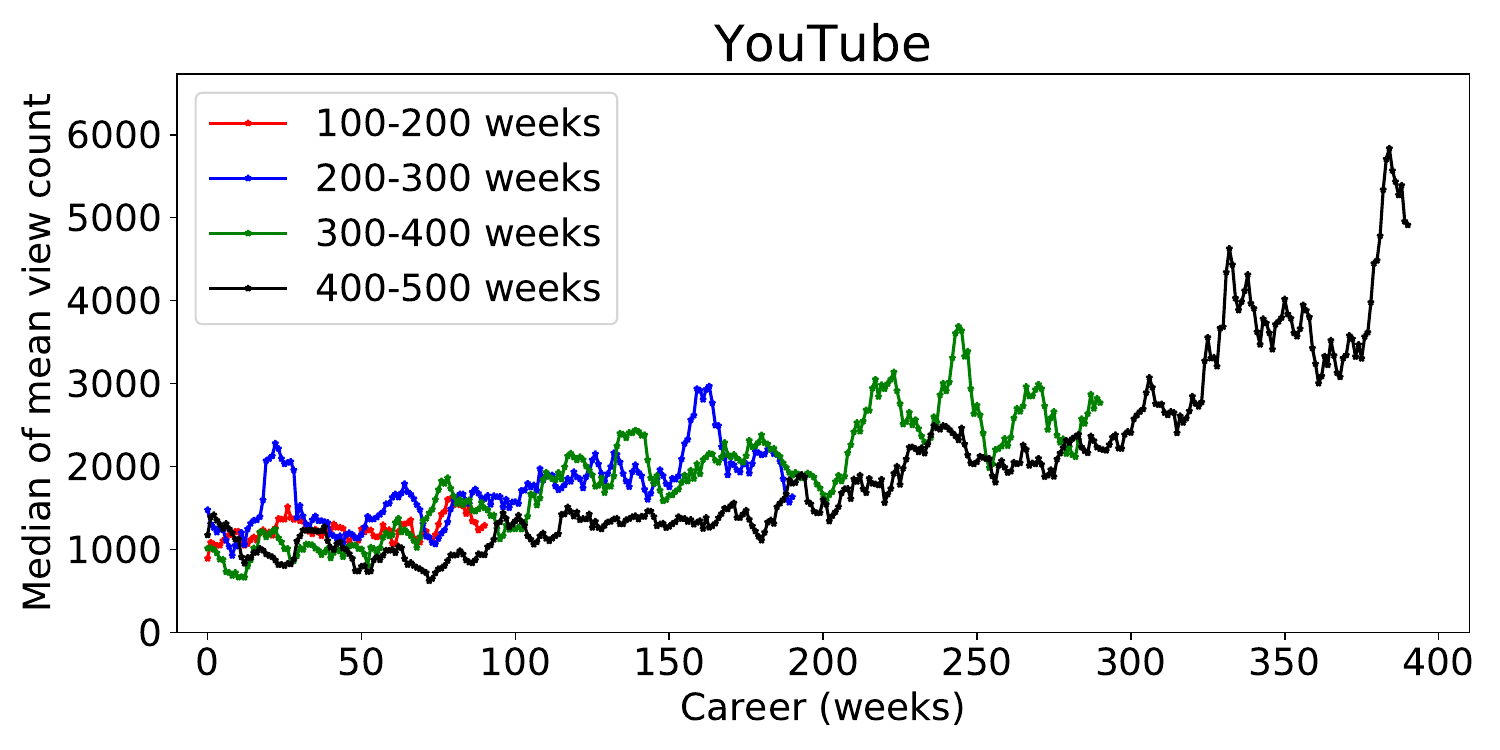}
\caption{
Evolution of retweet counts in user careers
(replication of \Figref{fig:rt_evo_raw} for all six datasets described in Appendix~\ref{sec:Description of datasets}).
The top left plot (\political) is identical to \Figref{fig:rt_evo_raw}.
}
\label{fig:rt_evo_raw_6_datasets}
\end{figure*}

\Figref{fig:rt_evo_raw_6_datasets} plots the evolution of retweet counts in user careers for all six datasets described in Appendix~\ref{sec:Description of datasets}, thus replicating the analysis of \Figref{fig:rt_evo_raw}.
As we see, the overall curve shape is similar across datasets.

\begin{figure*}
\centering
\includegraphics[width=0.33\textwidth]{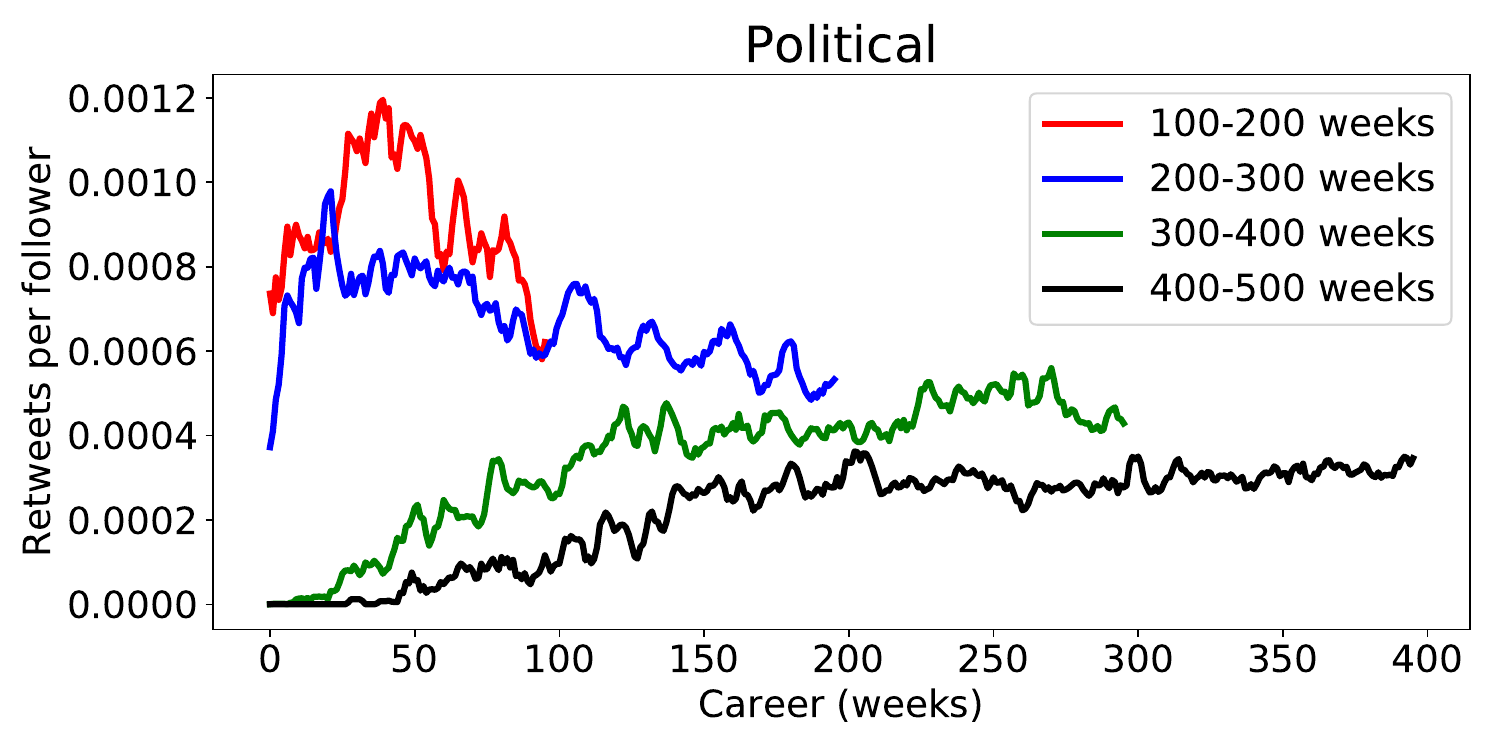}
\includegraphics[width=0.33\textwidth]{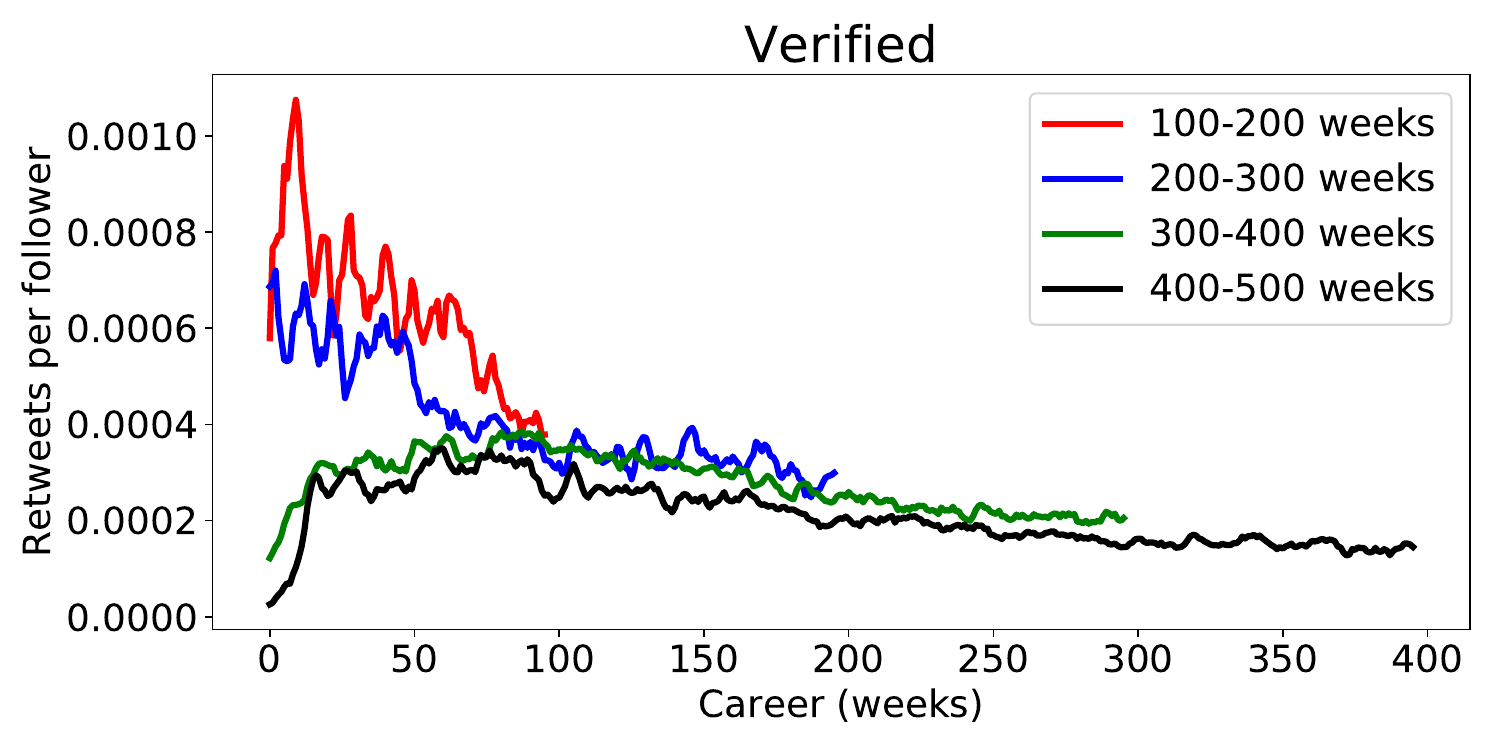}
\includegraphics[width=0.33\textwidth]{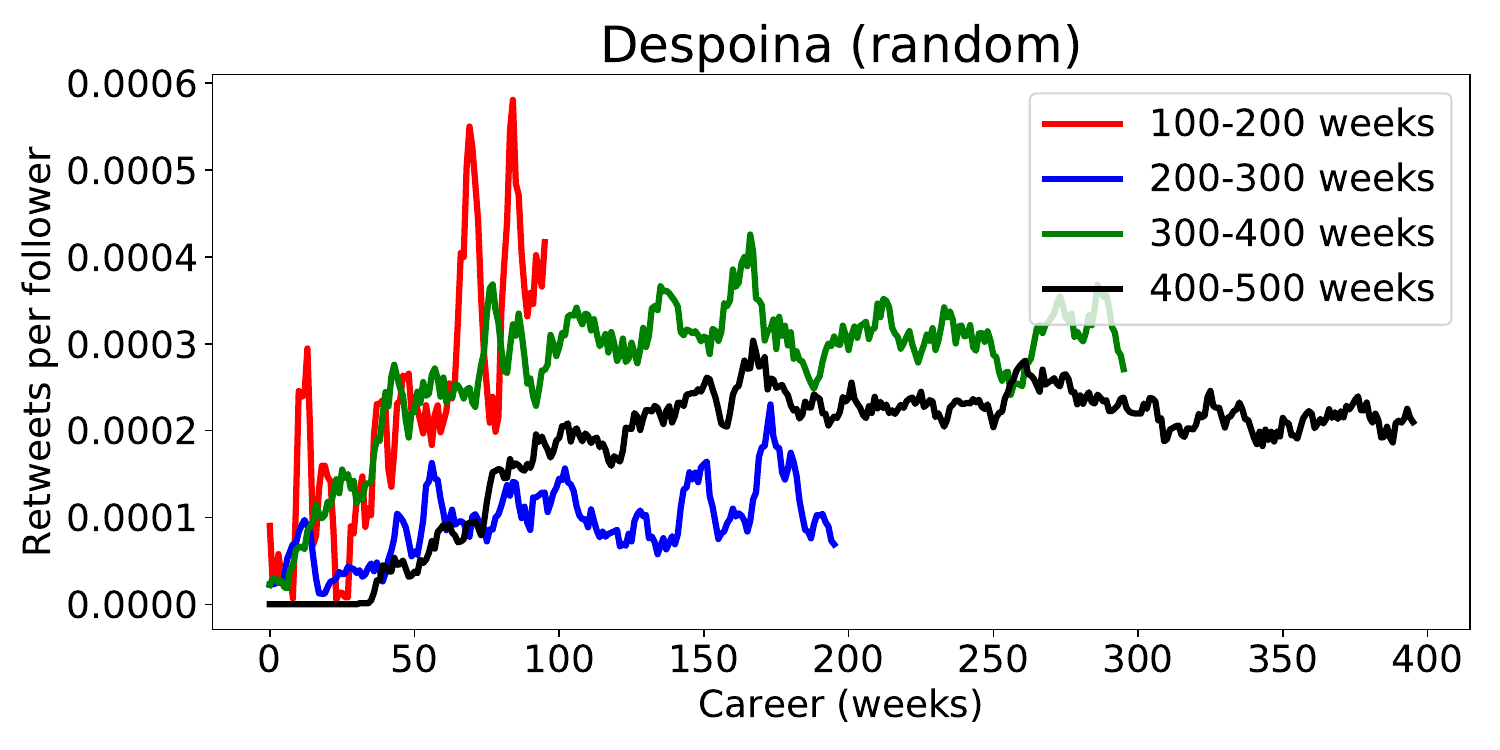}
\includegraphics[width=0.33\textwidth]{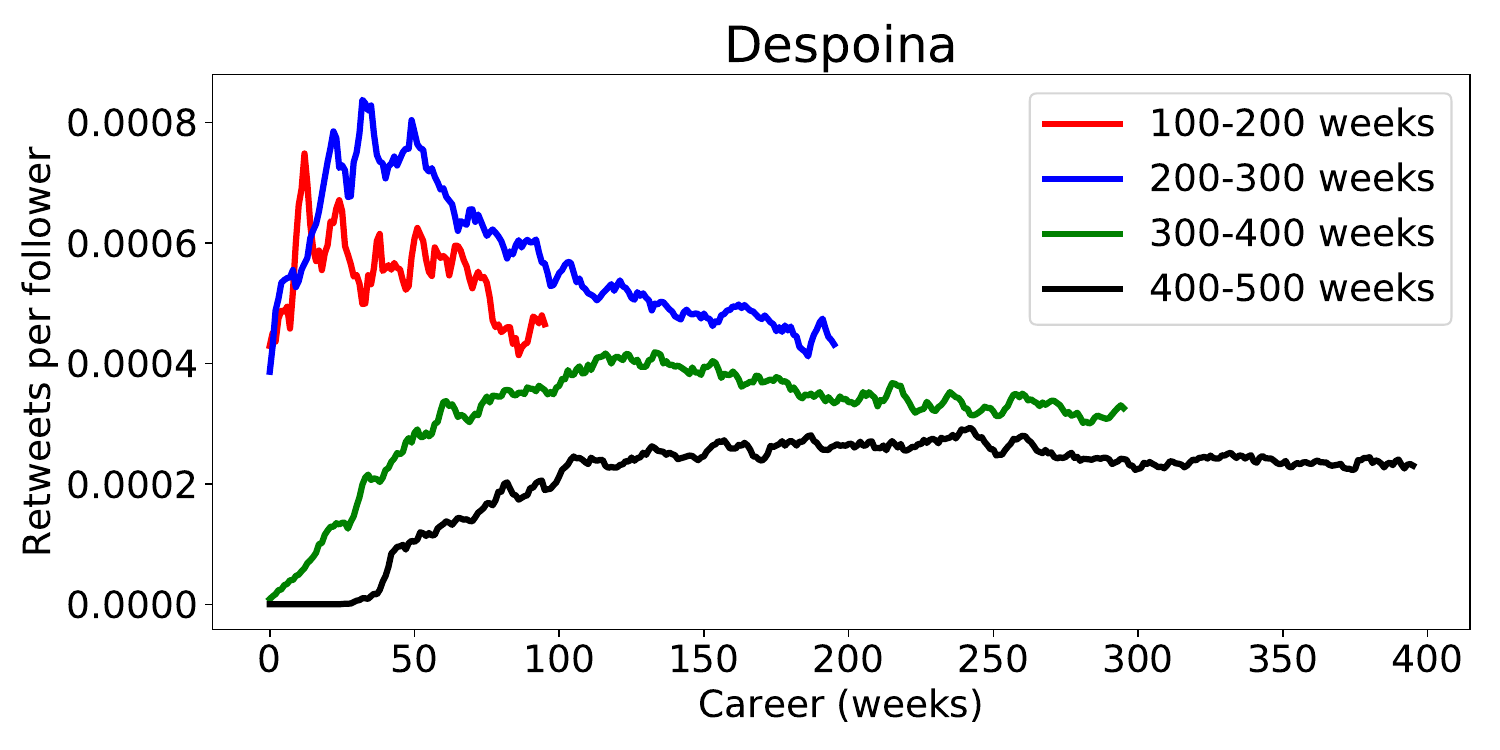}
\includegraphics[width=0.33\textwidth]{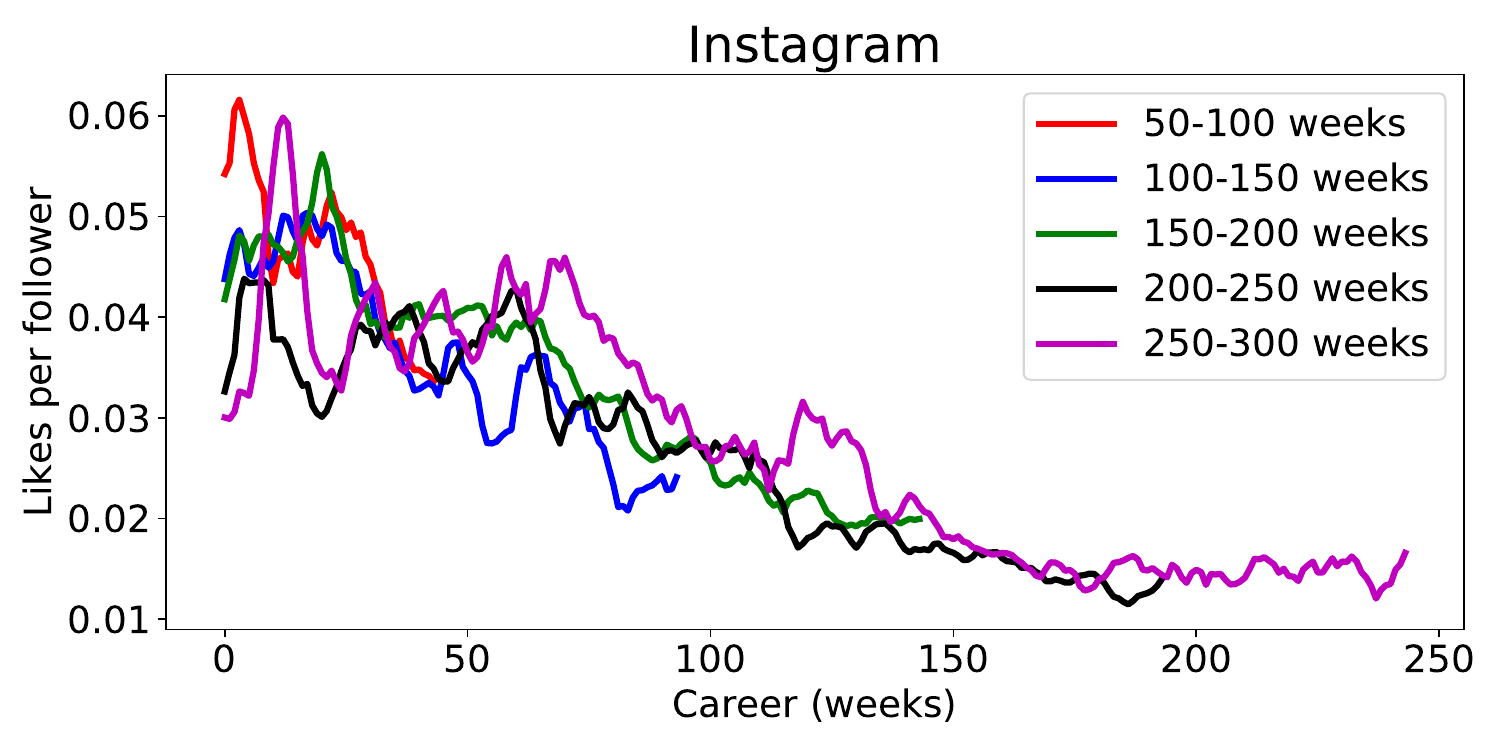}
\includegraphics[width=0.33\textwidth]{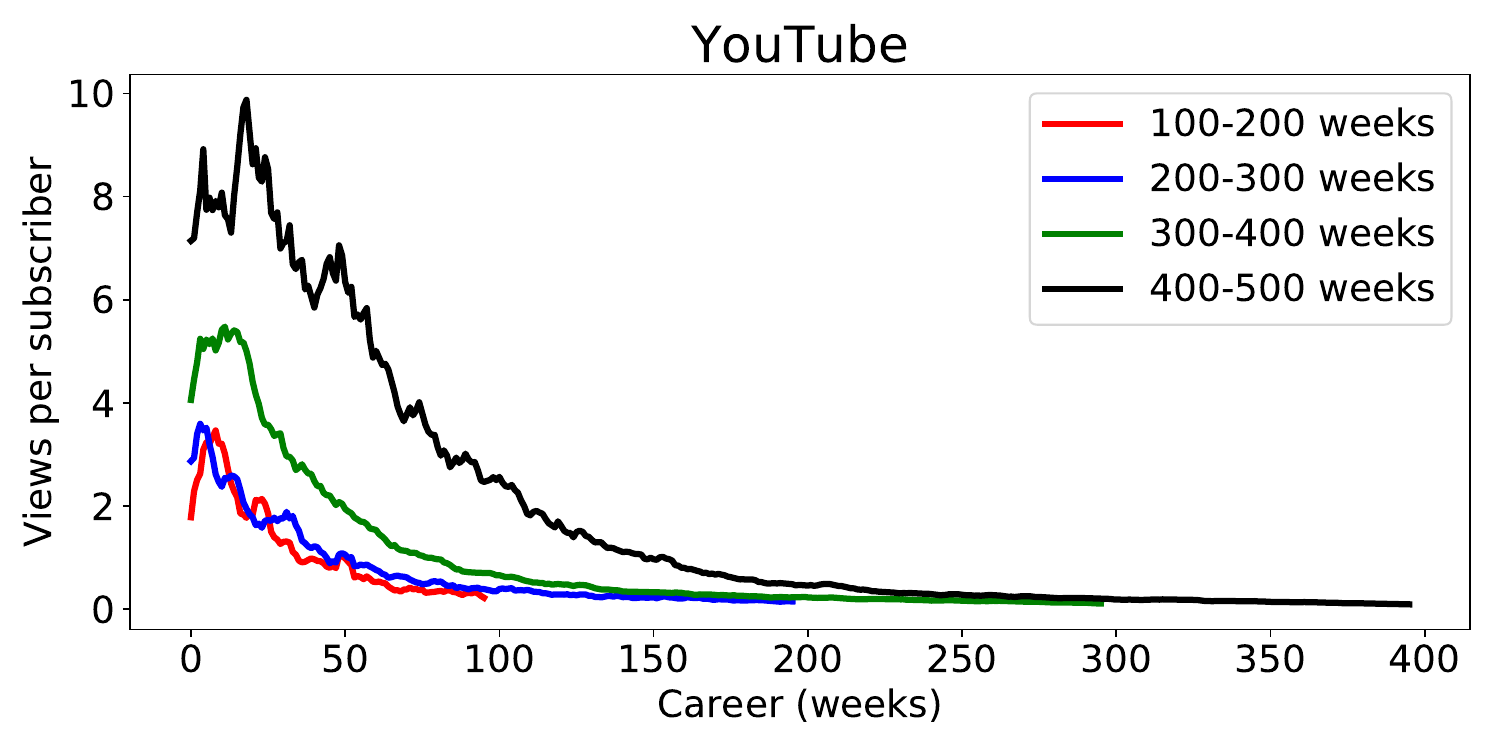}
\caption{
Evolution of per-follower retweet rates in user careers
(replication of \Figref{fig:rt_evo_norm} for all six datasets described in Appendix~\ref{sec:Description of datasets}).
The top left plot (\political) is identical to \Figref{fig:rt_evo_norm}.
}
\label{fig:rt_evo_norm_6_datasets}
\end{figure*}

Similarly, \Figref{fig:rt_evo_norm_6_datasets} plots the evolution of per-follower retweet rates in user careers for all six datasets, thus replicating the analysis of \Figref{fig:rt_evo_norm}.
Again, the characteristic curve shape is preserved: most curves have a non-monotonic shape with an internal ``peak age''.

\begin{figure}
\centering
\includegraphics[width=\columnwidth]{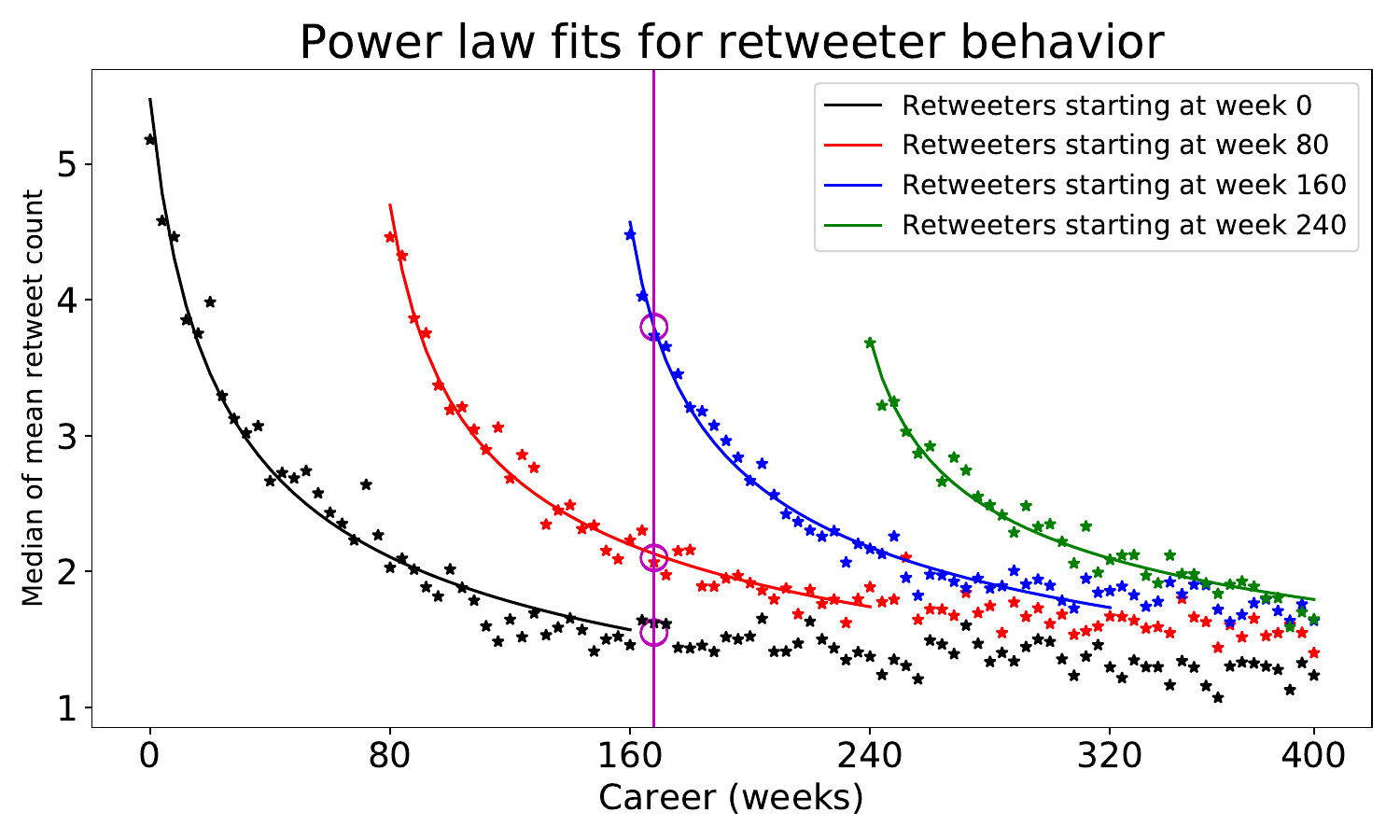}
\caption{
Power-law decay in followers' retweet rates for the \verified dataset, thus replicating \Figref{fig:assumption2} (which was based on the \political dataset).
}
\label{fig:powerlaw_fit}
\end{figure}

\Figref{fig:powerlaw_fit} plots the power-law decay in followers' retweet rates for the \verified dataset, thus replicating the analysis of \Figref{fig:assumption2} (which was based on the \political dataset).

\setcounter{secnumdepth}{0}

\bibliographystyle{aaai21}
\bibliography{biblio}

\begin{thebibliography}{23}
\providecommand{\natexlab}[1]{#1}
\providecommand{\url}[1]{\texttt{#1}}
\providecommand{\urlprefix}{URL }
\expandafter\ifx\csname urlstyle\endcsname\relax
  \providecommand{\doi}[1]{doi:\discretionary{}{}{}#1}\else
  \providecommand{\doi}{doi:\discretionary{}{}{}\begingroup
  \urlstyle{rm}\Url}\fi

\bibitem[{Abisheva et~al.(2014)Abisheva, Garimella, Garcia, and
  Weber}]{abisheva2014watches}
Abisheva, A.; Garimella, K.; Garcia, D.; and Weber, I. 2014.
\newblock Who watches (and shares) what on YouYube? And when? Using Twitter to
  understand YouTube viewership.
\newblock In \emph{WSDM}.

\bibitem[{Adams(1946)}]{adams1946age}
Adams, C.~W. 1946.
\newblock The age at which scientists do their best work.
\newblock \emph{Isis} 36(3/4): 166--169.

\bibitem[{Antonakaki, Ioannidis, and
  Fragopoulou(2018)}]{antonakaki2018utilizing}
Antonakaki, D.; Ioannidis, S.; and Fragopoulou, P. 2018.
\newblock Utilizing the average node degree to assess the temporal growth rate
  of Twitter.
\newblock \emph{Social Network Analysis and Mining} 8(1): 12--20.

\bibitem[{Barrick and Mount(1991)}]{barrick1991big}
Barrick, M.~R.; and Mount, M.~K. 1991.
\newblock The big five personality dimensions and job performance: A
  meta-analysis.
\newblock \emph{Personnel Psychology} 44(1): 1--26.

\bibitem[{Danescu-Niculescu-Mizil et~al.(2013)Danescu-Niculescu-Mizil, West,
  Jurafsky, Leskovec, and Potts}]{danescu2013no}
Danescu-Niculescu-Mizil, C.; West, R.; Jurafsky, D.; Leskovec, J.; and Potts,
  C. 2013.
\newblock No country for old members: User lifecycle and linguistic change in
  online communities.
\newblock In \emph{WWW}.

\bibitem[{DeNisi and Stevens(1981)}]{denisi1981profiles}
DeNisi, A.~S.; and Stevens, G.~E. 1981.
\newblock Profiles of performance, performance evaluations, and personnel
  decisions.
\newblock \emph{Academy of Management Journal} 24(3): 592--602.

\bibitem[{Dennis(1966)}]{dennis1966creative}
Dennis, W. 1966.
\newblock Creative productivity between the ages of 20 and 80 years.
\newblock \emph{Journal of Gerontology} 21(1): 1--8.

\bibitem[{Figueiredo and Almeida(2011)}]{figueiredo2011tube}
Figueiredo, F.; and Almeida, J. 2011.
\newblock The tube over time: Characterizing popularity growth of YouTube
  videos.
\newblock In \emph{WSDM}.

\bibitem[{Garimella et~al.(2018)Garimella, Morales, Gionis, and
  Mathioudakis}]{garimella2018political}
Garimella, K.; Morales, G. D.~F.; Gionis, A.; and Mathioudakis, M. 2018.
\newblock Political discourse on social media: Echo chambers, gatekeepers, and
  the price of bipartisanship.
\newblock In \emph{WWW}.

\bibitem[{Garimella and West(2019)}]{garimella2019hot}
Garimella, K.; and West, R. 2019.
\newblock Hot streaks on social media.
\newblock In \emph{ICWSM}.

\bibitem[{Goel et~al.(2016)Goel, Anderson, Hofman, and
  Watts}]{goel2016structural}
Goel, S.; Anderson, A.; Hofman, J.; and Watts, D.~J. 2016.
\newblock The structural virality of online diffusion.
\newblock \emph{Management Science} 62(1).

\bibitem[{Kupavskii et~al.(2012)Kupavskii, Ostroumova, Umnov, Usachev,
  Serdyukov, Gusev, and Kustarev}]{kupavskii2012prediction}
Kupavskii, A.; Ostroumova, L.; Umnov, A.; Usachev, S.; Serdyukov, P.; Gusev,
  G.; and Kustarev, A. 2012.
\newblock Prediction of retweet cascade size over time.
\newblock In \emph{CIKM}.

\bibitem[{Lehman(1953)}]{lehman2017age}
Lehman, H.~C. 1953.
\newblock \emph{Age and Achievement}.
\newblock Princeton University Press.

\bibitem[{Liu et~al.(2018)Liu, Wang, Sinatra, Giles, Song, and
  Wang}]{liu2018hot}
Liu, L.; Wang, Y.; Sinatra, R.; Giles, C.~L.; Song, C.; and Wang, D. 2018.
\newblock Hot streaks in artistic, cultural, and scientific careers.
\newblock \emph{Nature} 559(7714): 396.

\bibitem[{Martin et~al.(2016)Martin, Hofman, Sharma, Anderson, and
  Watts}]{martin2016exploring}
Martin, T.; Hofman, J.~M.; Sharma, A.; Anderson, A.; and Watts, D.~J. 2016.
\newblock Exploring limits to prediction in complex social systems.
\newblock In \emph{WWW}.

\bibitem[{Meeder et~al.(2011)}]{meeder2011we}
Meeder, B.; et~al. 2011.
\newblock We know who you followed last summer: Inferring social link creation
  times in Twitter.
\newblock In \emph{WWW}.

\bibitem[{Myers and Leskovec(2014)}]{myers2014bursty}
Myers, S.~A.; and Leskovec, J. 2014.
\newblock The bursty dynamics of the {T}witter information network.
\newblock In \emph{WWW}.

\bibitem[{Radicchi and Vespignani(2009)}]{radicchi2009diffusion}
Radicchi, F.; and Vespignani, A. 2009.
\newblock Diffusion of scientific credits and the ranking of scientists.
\newblock \emph{Physical Review E} 80: 056103.

\bibitem[{Sinatra et~al.(2016)Sinatra, Wang, Deville, Song, and
  Barab{\'a}si}]{sinatra2016quantifying}
Sinatra, R.; Wang, D.; Deville, P.; Song, C.; and Barab{\'a}si, A.-L. 2016.
\newblock Quantifying the evolution of individual scientific impact.
\newblock \emph{Science} 354(6312).

\bibitem[{Suh et~al.(2010)Suh, Hong, Pirolli, and Chi}]{suh2010want}
Suh, B.; Hong, L.; Pirolli, P.; and Chi, E.~H. 2010.
\newblock Want to be retweeted? Large scale analytics on factors impacting
  retweet in Twitter network.
\newblock In \emph{SocialCom}.

\bibitem[{Vilenchik(2018)}]{vilenchik2018million}
Vilenchik, D. 2018.
\newblock The million tweets fallacy: Activity and feedback are uncorrelated.
\newblock In \emph{ICWSM}.

\bibitem[{Yucesoy and Barab{\'a}si(2016)}]{yucesoy2016untangling}
Yucesoy, B.; and Barab{\'a}si, A.-L. 2016.
\newblock Untangling performance from success.
\newblock \emph{EPJ Data Science} 5(1): 17.

\bibitem[{Zhang et~al.(2018)Zhang, Chen, Zhou, Li, and Luo}]{zhang2018become}
Zhang, Z.; Chen, T.; Zhou, Z.; Li, J.; and Luo, J. 2018.
\newblock How to become Instagram famous: Post popularity prediction with
  dual-attention.
\newblock In \emph{BigData}.

\end{thebibliography}

\end{document}